# Large-Area Photonic Lift-off Process for Flexible Thin-Film Transistors

*Adam M. Weidling[1], Vikram S. Turkani[2], Vahid Akhavan[2], Kurt A. Schroder[2]\*, Sarah L. Swisher[1]\**

[1]Department of Electrical and Computer Engineering, University of Minnesota Twin Cities
4-174 Keller Hall, 200 Union Street S.E., Minneapolis, MN 55455, USA
[2]NovaCentrix, 400 Parker Drive, Suite 1110, Austin, TX 78728, USA
\* Co-corresponding authors. Email: kurt.schroder@novacentrix.com, sswisher@umn.edu

## ABSTRACT

Fabricating high-performance and/or high-density flexible electronics on plastic substrates is often limited by the poor dimensional stability of polymer substrates. This can be mitigated by using glass carriers during fabrication, but removing the plastic substrate from a large-area carrier without damaging the electronics remains challenging. Here we present a large-area photonic lift-off (PLO) process to rapidly separate polymer films from rigid carriers. PLO uses a 150 µs pulse of broadband light from flashlamps to lift off functional thin films from a glass carrier substrate coated with a light-absorber layer (LAL). A 3D finite element model indicates that the polymer/LAL interface reaches 865 ˚C during PLO, but the top surface of the PI reaches only 118 ˚C. To demonstrate the feasibility of this process in the production of flexible electronics, an array of indium zinc oxide (IZO) thin-film transistors (TFTs) was fabricated on a polyimide substrate and then photonically lifted off from the glass carrier. The TFT mobility was 3.15 $cm^2V^{-1}s^{-1}$ before and after PLO, indicating no significant change during PLO. The flexible TFTs were mechanically robust, with no reduction in mobility while bent. The PLO process can offer unmatched high-throughput solutions in large-area flexible electronics production.

# 1 INTRODUCTION

As the complexity of flexible electronic devices such as flexible displays and sensors has increased, new plastic-compatible manufacturing methods are becoming necessary to produce high-yield, low-defect flexible circuits. While solution-processed methods such as printing can be used to fabricate circuits on low-temperature substrates [1], printing is not always a feasible option to realize complex, high-density circuits with small devices. In particular, high-resolution display applications typically require single-micron feature sizes [2,3] and excellent uniformity over large-area substrates. For these applications, more precise fabrication techniques such as photolithography are required. However, due to the poor dimensional stability of polymers, even low-temperature processes such as soft bakes during lithography can alter the substrate dimensions. As the feature size decreases, even micron-scale changes in the substrate dimensions can introduce errors in subsequent alignment steps [4,5]. One solution to this problem is transfer printing [6], in which materials are fabricated on one substrate and then transferred to a receiver substrate. However, transfer printing can suffer from low yield without the use of an adhesion layer or careful tuning of the substrates' surface energies. Alternatively, the problem of dimensional stability can be overcome by attaching the polymer film to a rigid carrier during device fabrication, then releasing the polymer substrate with the fully fabricated device layer on top. For this reason, the separation of polymer films from rigid carriers remains a crucial step in creating next-generation flexible electronics.

A variety of methods have been investigated to release a plastic substrate from a rigid carrier. Laser lift-off (LLO) utilizing excimer lasers has long been the method used in industry to delaminate polymer films from the rigid carriers that are required during electronics fabrication [7–17]. The mechanism of lift-off is based on the selective ablation of the polymer film at the interface between the polymer and the rigid carrier. The interaction between polymers and lasers has been extensively studied [18–26], and these studies led to the LLO technique. In LLO, the laser is scanned across the back of the glass substrate and ablation is achieved by absorption occurring in the polymer substrate itself, as polymers typically absorb wavelengths in the ultraviolet (UV) region of the spectrum. The absorbed laser energy heats the polymer and causes ablation at the glass/polymer interface. Excimer LLO has several attractive properties such as the ability to maintain local radiation and achieve low global temperatures [17]. However, expensive laser systems are required, and the lift-off process is sensitive to variations in the laser parameters. Power fluctuations and non-uniform rastering of the laser system can cause defects and reduce yield [14]. LLO is also a time-intensive rastering process, which translates to higher production costs when working on large-area substrates [27].

To overcome the challenges of LLO, alternative delamination techniques have been investigated. Temporary bonding layers such as tapes and adhesives have been used to attach the plastic film to the carrier during fabrication, but even some of the bonding materials themselves produce small dimensional changes [28–33]. Many also require UV or thermal energy [30,34–37], solvents [34], or mechanical force to delaminate the sample [30,38,39]. Inorganic separation layers such as oxidized tungsten are more stable than adhesives at high processing temperatures, but still require substantial delamination force [40]. Thermal ablation from joule heating has also been demonstrated [41], but is not widely used.

In this study, we present a non-laser photonic lift-off (PLO) process to overcome the challenges presented by the previous techniques used to separate polymer films from rigid carriers. PLO uses broadband light (200 nm – 1100 nm) from a flashlamp to lift off a polymer film from a rigid carrier coated with a light-absorbing layer (LAL), where the polymer film may contain functional devices [42]. Preliminary results using flashlamps instead of lasers for polymer lift-off were reported recently [43,44], but no active devices or TFTs were included. Compared to those results, PLO in this work utilizes a simple single-layer metal absorber rather than a 4-layer structure, and an ultra-fast pulse duration that is only 2~5% of the previously reported pulse length. Compared to a raster process such as LLO, the PLO process is advantageous as it can enable large area lift-off (150 mm x 75 mm) in one flash of light lasting between 100- 200 µs, thus enabling higher throughput. Furthermore, the presence of the LAL prevents direct illumination of the polymer substrate during lift-off and facilitates cleaner lift-off without polymer ashing, which is a common concern with LLO. The presence of the LAL makes PLO a polymer-agnostic process enabling lift-off of a wide range of polymers.

Here we present the first reported fabrication of flexible thin-film transistors (TFTs) using PLO. Polyimide with a low coefficient of thermal expansion (CTE) was used to minimize residual film stress. Indium zinc oxide (IZO) TFTs were fabricated on the polyimide films and subsequently released from the carrier using PLO. The released devices demonstrated no significant changes after the PLO process, yielding a mobility of ~3 $cm^2V^{-1}s^{-1}$ before and after lift-off. The flexible TFTs were robust to bending stress, showing no notable decrease in mobility while bent with a 10 mm radius of curvature. When combined with the ability to successfully separate the PI from the carrier, the low-CTE PI creates a robust fabrication platform for flexible electronics that allows conventional patterning and deposition techniques to be used on a plastic substrate while maintaining dimensional stability up to 380 °C. A 3D finite element model was developed to validate the thermal response in the substrate during lift-off. The modeled temperature in the polyimide film reaches 865 °C at the ablation layer but stays below 120 °C at the device layer, which is low enough to avoid damage or induce chemical changes in the thin-film transistors. This work demonstrates the feasibility of PLO to successfully release polyimide substrates containing flexible electronic devices from glass carriers after withstanding conventional lithographic processing methods.

## 2  RESULTS AND DISCUSSION

### 2.1  Photonic Lift-off Process

Figure 1a shows a typical material stack used in PLO to lift off a polymer carrying electronic devices. First, a glass carrier substrate is coated with a thin metal light-absorber layer (LAL), typically 250 nm of sputtered TiW (10:90). A polymer coating is then applied to the LAL, and finally, the electronic structures are fabricated on the surface of the polymer film. The PLO process involves irradiating the LAL through the glass with a broadband (200 nm – 1100 nm) light pulse using the PulseForge photonic curing system (Figure 1a). The PulseForge generates an intense pulse of light (output up to 45 kW/cm$^2$) with a 150 μs pulse length, providing adequate heating over a short time period to lift off the polymer. Figure 1b shows the simulated thermal response of the LAL during the PLO process, illustrating that the temperature at the LAL exceeds 850 °C in less than 0.5 ms, but the top surface of the polyimide stays below 120 °C. A flexible array of TFTs on a polyimide substrate that was delaminated using photonic lift-off is shown in Figure 1c.

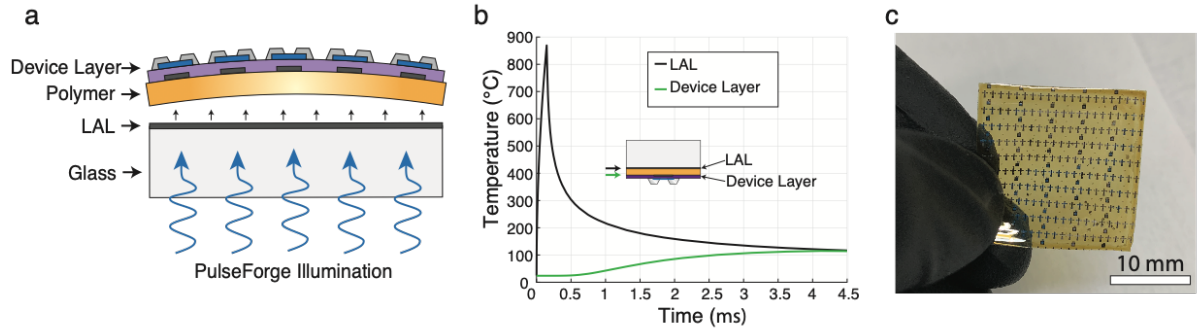

Figure 1: (a) Typical material stack for the photonic lift-off (PLO) process, showing illumination through the glass to the light-absorber layer (LAL). (b) Simulated thermal response during PLO at the polymer/LAL interface and on the surface of the polyimide where the device layer would reside. (c) Flexible TFT array on polyimide lifted off using PLO.

The metal LAL is the key enabler of the PLO process. Its primary purpose is to convert the absorbed optical energy into the thermal energy needed to facilitate the ablation of the polymer at the metal interface, enabling lift-off of the polymer. Its secondary purpose is to block the transmission of the broadband light into the polymer and device layers. Thus, the design requirements for the LAL are that the optical transmittance should be nearly zero and the reflectance should be minimized, and the absorbance maximized. Not only are the optical parameters important, but the adhesion between the LAL and the glass is also crucial. The metal must also have sufficient thermal stability to withstand high temperatures without cracking. In general, these design guidelines are similar to those we have previously discussed for photonic curing of IZO-based TFTs [45]. In this work, we utilized TiW (10:90) for the LAL due to excellent adhesion to the glass, high optical absorption, and thermal robustness. The coefficient of thermal expansion also closely matches that of the glass substrate (4~4.5 ppm°C$^{-1}$ and 3.2 ppm°C$^{-1}$, respectively). Furthermore, it was observed that cleaning the glass prior to deposition of the absorber layer was an important factor in the overall success of PLO. If improperly cleaned, the adhesion between the LAL and the glass can be a failure point causing the metal to delaminate during PLO.

The optical properties of the LAL were characterized using UV-Vis spectroscopy to determine its spectral absorbance. Using an integrating sphere, the reflectance and transmittance were measured, and the absorbance was calculated ($A = 100 - T - R$). Optical measurements (Figure 2) with light incident on the back of the glass substrate reveal that the absorber layer coating has ≈0% transmission throughout the mid-UV to the near-IR region, as desired. These measurements also confirm the high absorbance in the TiW LAL, which is essential for efficient heating [45]. The absorbance of the LAL is >55% in the wavelength range corresponding to the dominant emission from the PulseForge lamp ($\lambda$ = 400~600 nm, Figure S1), providing sufficient absorption in the LAL for photonic lift-off.

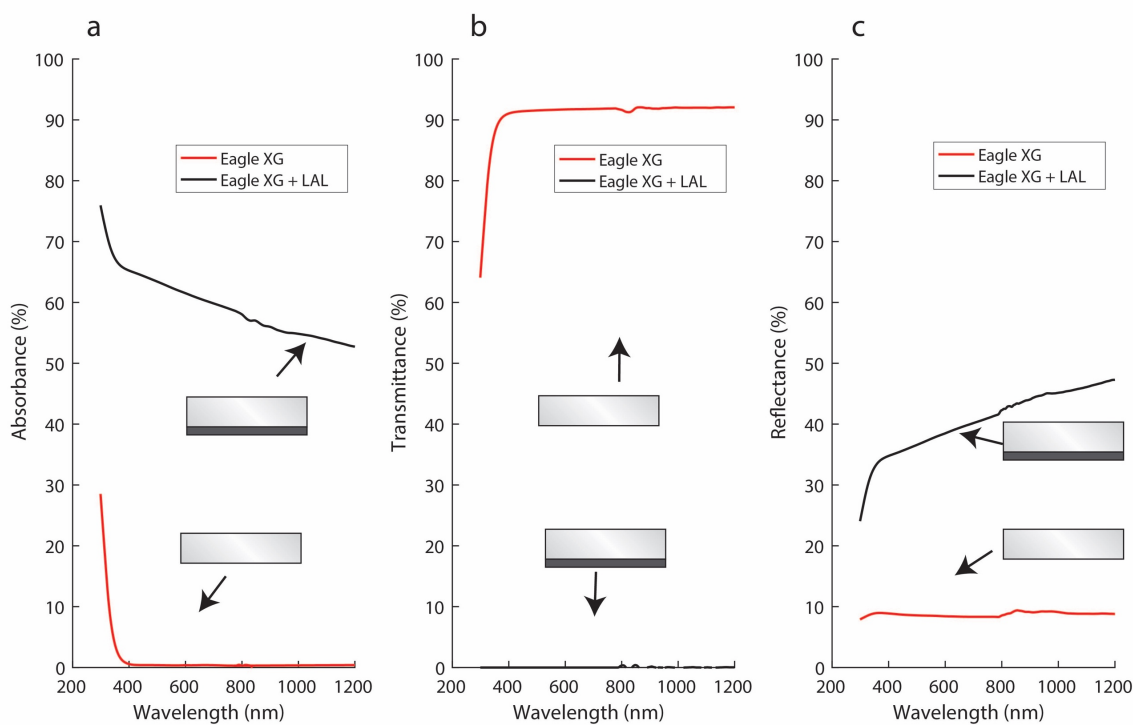

Figure 2: UV-Vis (a) absorbance, (b) transmittance, and (c) reflectance spectra for Eagle XG glass with and without the 250 nm TiW light absorber layer (LAL).

Because the LAL protects the polymer layer from direct illumination, it makes the PLO a polymer-agnostic process. To demonstrate this, a variety of both thermoset and thermoplastic polymers were lifted off from the LAL-coated glass carrier using the PLO process. Figure S2 shows successful PLO of a thermoset polymer such as polyimide (PI 2525 HD Microsystems), clear polyimide (KOLON CPI™), and a thermoplastic polymer such as polyurethane (Lubrizol Sancure™ 1511).

## 2.2 Finite Element Modeling

The PLO process involves converting the incident optical energy on the LAL into thermal energy to ablate a polymer. Thus, it is imperative to study the thermal response at the LAL-polymer interface during photonic lift-off. To estimate the thermal response, a 3D finite element analysis model for the PLO process was developed using COMSOL® Multiphysics. The geometry in the model was defined to match the stack used in this process (700 µm Eagle XG glass, 250 nm TiW, and 30 µm PI). The 3D heat conduction equation used to solve for the thermal gradients is shown in Equation 1.

$$\rho c_p \frac{\partial T}{\partial t} + \nabla \cdot q = 0 \qquad (1)$$

Here $\rho$ is the material density, $c_p$ is the specific heat capacity, T is the temperature, $q = -\kappa \nabla T$ is the local heat flux density and $\kappa$ is the thermal conductivity. The density, heat capacity, and thermal conductivities used in the simulation are shown in Table 1.

| Material | Density [gcm$^{-3}$] | Heat Capacity [Jkg$^{-1}$C$^{-1}$] | Thermal Conductivity [Wm$^{-1}$K$^{-1}$] |
| --- | --- | --- | --- |
| TiW | 19.25 | 134 | 160 |
| Eagle XG | 3.28 | 1100 | 1.2 |
| Polyimide (2611) | 1.4 | 1009 | 0.12 |

Table1 : Material parameters used to create a 3D model of the thermal response during photonic lift-off.

The heat flux incident on the LAL was modeled using the deposited beam power physics module in COMSOL Multiphysics. The energy incident on the LAL was determined by measuring the radiant exposure of the lamp irradiating at typical lift-off conditions (950 V and 150 μs) on the bolometer, through the Eagle XG glass (Figure S3). The bolometer recorded an average radiant exposure of $3.26 \pm 0.05$ JCm$^{-2}$. For comparison, the radiant exposure measured directly by the bolometer (without the glass) was $4.51 \pm 0.1$ Jcm$^{-2}$. The radiant exposure measured through the glass was multiplied by the absorbance of the LAL obtained from UV-Vis spectroscopy. The product represents the energy absorbed in the LAL, which was 1.98 Jcm$^{-2}$.

A convective heat flux was defined at all boundaries using a convective heat transfer coefficient of 15 Wm$^{-1}$K$^{-1}$ to model the interaction of the sample with air because the sample is typically suspended during the PLO process. A time-dependent solution to Equation 1 was obtained and the resulting thermal profiles as a function of time are shown in Figure 3. Figure 3a shows temperature as a function of depth where 0 μm is the top surface of the PI where the thin-film electronics would reside, and 30 μm is the LAL-PI interface. From 30 μm to 730 μm is the bulk of the glass carrier. Note that the depth axis in Figure 3a has been truncated at 300 μm because beyond that point the glass remains at room temperature. The maximum temperature of 865 °C is achieved at the LAL-PI interface at 150 μs, which corresponds to the end of the PLO light pulse. It is important to note that the surface of the PI where the devices are fabricated does not see this exceedingly high temperature. As time progresses, since the thermal conductivity of the glass (1.2 Wm$^{-1}$K$^{-1}$) is higher than that of the PI (0.12 Wm$^{-1}$K$^{-1}$), the heat conducts into the glass first whereas the heat conduction into PI is slower. Figure 3b shows the thermal gradient through a cross-section of the entire device stack at various time intervals, demonstrating that the heat is confined to a very thin region to ablate the PI and that the temperature rapidly drops below the maximum working temperature of the PI. The simulation shows that the maximum temperature obtained at the surface of the PI during lift-off is 118 ˚C. In reality, after the peak temperature is reached, the polyimide is no longer in perfect contact with the LAL due to the ablation, so we hypothesize that this simulation is an overestimate and that the actual temperature at the surface of the PI is lower. In this way, the PLO process is self-limiting. The temperature obtained on the surface of the PI is well within the operating limits of standard polyimide films and is no higher than a typical soft bake temperature during lithography. Based on these results, the PLO process is not expected to damage electronic devices on the surface of the polyimide.

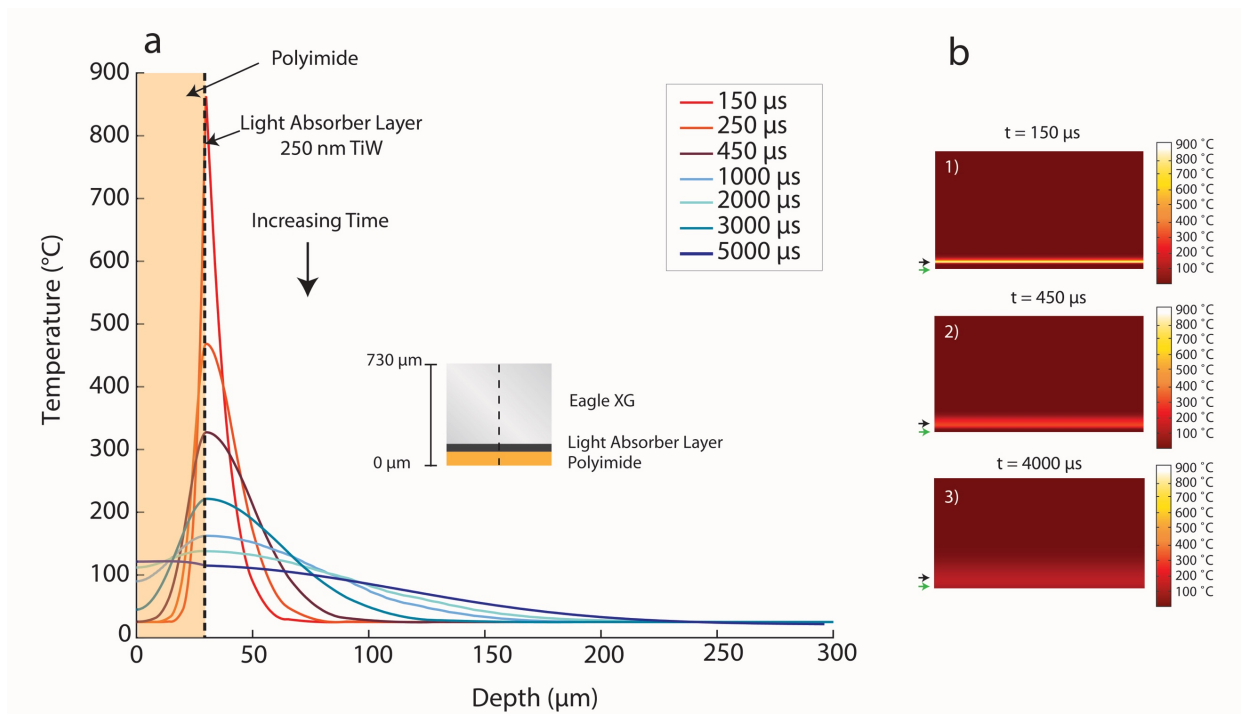

Figure 3: a) Thermal response as a function of depth at various time intervals. The depth plots were truncated at 300 µm into the glass because no temperature rise was present beyond that point. b) Heatmaps showing the temperature gradient across the entire 730 µm stack depth at various times, where the black and green arrows indicate the position of the LAL and the top surface of the polyimide, respectively. 1) Thermal cross-section when the maximum temperature is reached ($t$ = 150 µs), 2) thermal cross-section at 450 µs, and 3) thermal cross-section as sample reaches thermal equilibrium ($t$ = 4000 µs).

The low thermal conductivity of the PI ensures that the heat conduction to the top surface is minimized. However, as the PI thickness decreases, the heat does not have to penetrate as far to reach the devices, and the temperature experienced on the device surface of the PI increases. The peak temperature at the surface of the PI was simulated using a 950 V, 150 µs pulse using the same structure as shown in Figure 3, but with the PI film thicknesses ranging from 20 µm to 50 µm. As the film thickness increased, the peak temperature on the device surface of the PI decreased from 155 ˚C to 88 ˚C (Figure 4). This shows that the temperature on the PI surface is inversely related to the PI thickness. Thus, the resulting heat transfer to the devices depends on the polymer thickness and its thermal conductivity.

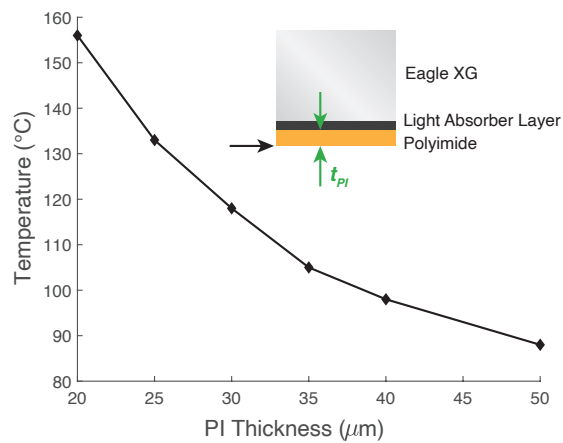

Figure 4: Simulated peak temperature at the surface of the polyimide (PI) as a function of the thickness of the PI layer for a 950 V, 150 µs photonic lift-off pulse. The inset shows the Eagle XG/TiW/PI material stack, with green arrows indicating the thickness of the PI layer and a black arrow indicating that the simulated temperature refers to the top surface of the polyimide.

## 2.3  Photonic Lift-off of TFTs on PI

Indium zinc oxide TFTs fabricated on a 30 µm thick polyimide layer (PI 2611) were subjected to the PLO process to separate the polyimide substrate from the LAL-coated glass carrier. PI 2611 is a low-CTE version of polyimide, which was important to minimize residual film stress after PLO. The sample was face down and suspended on glass microscope slides during the PLO process, such that the light passes through the glass carrier and is incident on LAL (Figure 5). Suspending the sample ensures that the active layer does not contact the platen and damage the thin-film devices. The released PI film containing flexible IZO TFTs is shown in Figure 5.

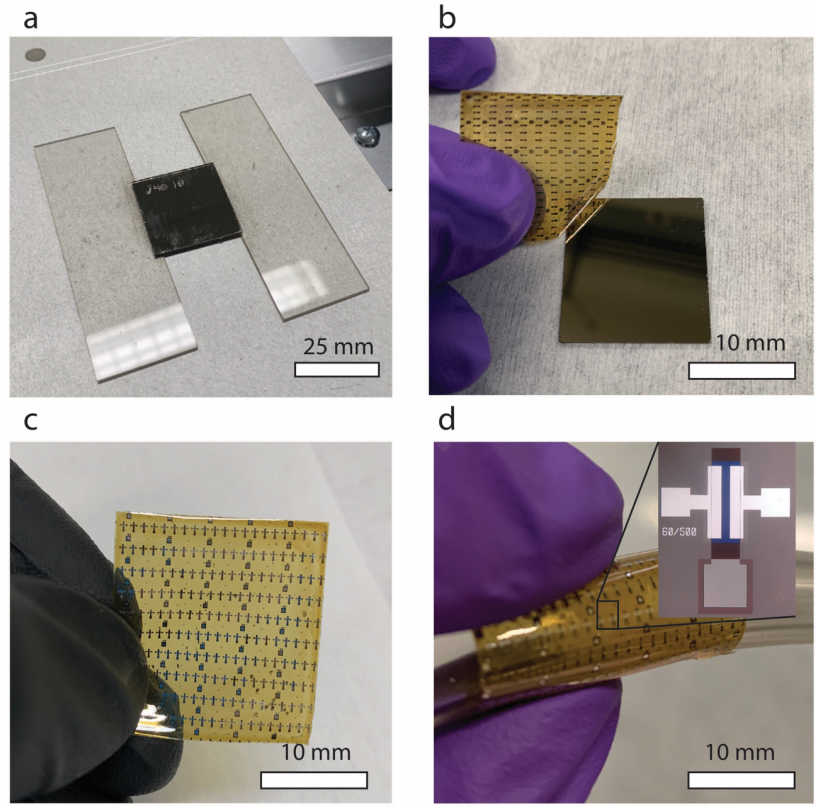

Figure 5: a) Device suspended on glass slides before PLO. b) Device being removed from the LAL after PLO. c) Flexible TFTs on polyimide after release from the carrier substrate. d) Device bent around a plastic tube and inset showing a W/L = 500/60 µm device on polyimide.

During PLO, the temperature in the LAL must exceed the decomposition temperature of the polymer to successfully achieve lift-off. The PLO process condition was optimized by varying the pulse duration from 130 to 160 µs at a fixed capacitor bank voltage of 950 V, where longer pulses lead to higher incident exposure and result in higher temperatures. Figure 6a shows the LAL-coated glass carrier and the PI substrates lifted off with varying pulse durations. Pulse duration and capacitor bank voltage are the two primary process variables in the PulseForge photonic curing tool, collectively contributing towards the energy density and incident power output of the system. Energy densities obtained as a result of varying pulse duration were correlated to the modeled maximum temperature at the LAL-PI interface (Figure 6b). The samples were successfully lifted off when the pulse duration was below 140 µs, but the released devices exhibited a large degree of PI curling, which indicates residual stress in the PI. These PI samples required a relatively higher peel force to separate them from the carrier after the PLO. This increase in the peel force indicates some residual adhesion between the PI and the carrier was still present after PLO. Devices lifted off with pulse lengths greater than 140 µs exhibited less curling and laid flat once released. Although the specified decomposition temperature of PI-2611 is 610 ˚C, we found it advantageous to use PLO pulses of at least 140 µs (corresponding to a peak temperature of ~750 ˚C) to ensure easy release from the carrier. We did not notice any soot formation on the LAL in any of these devices, which has been observed in previous studies [43]. As the pulse width increases the resulting temperature at the surface of the PI is likely to increase. Especially in thin PI films, this can result in damage to the devices, so pulse durations should be kept as short as possible without inducing curling or sacrificing lift-off yield.

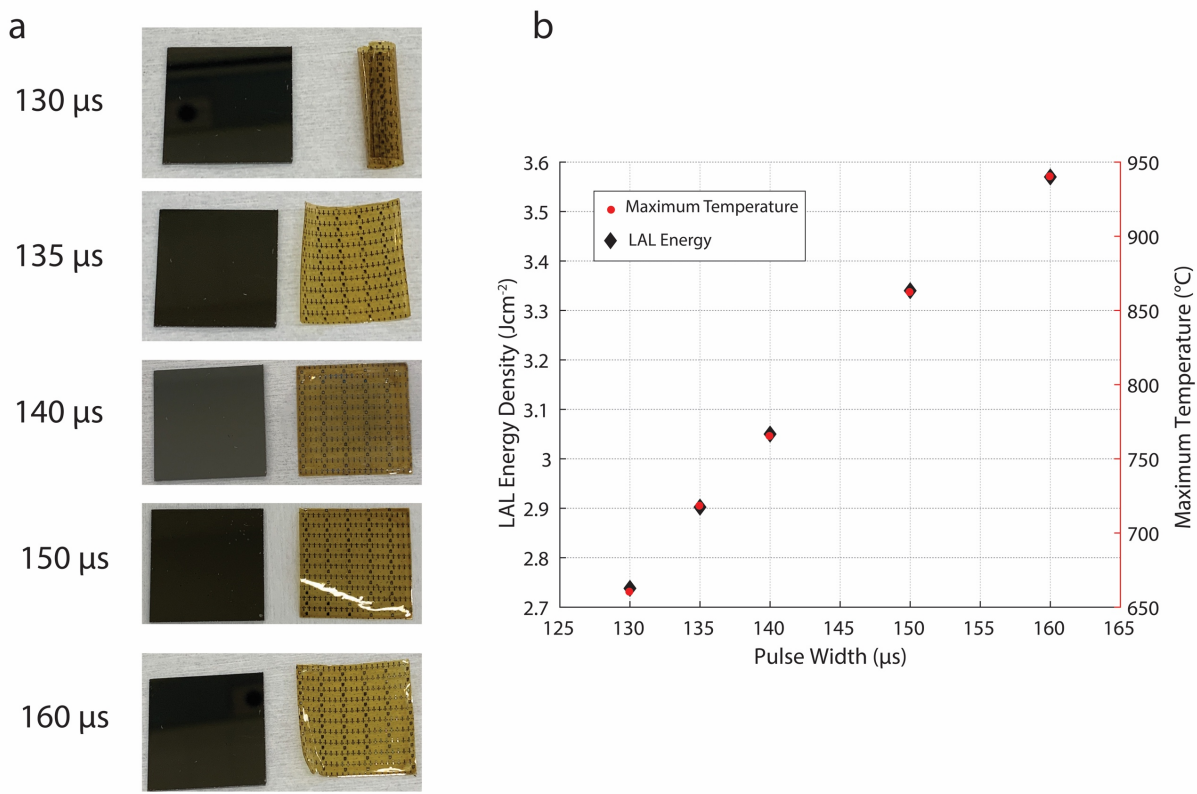

Figure 6: (a) Released samples that underwent PLO at 950 V with pulse widths ranging from 130 µs to 160 µs. (b) Energy incident on the LAL and the maximum simulated LAL temperature.

## 2.4 TFT Characterization

The IZO TFTs were characterized before and after the PI was released from the glass carrier using PLO. The TFT transfer and output curves were measured (Figure 7), and device metrics such as mobility, subthreshold slope, and leakage current were extracted. Prior to PLO while the polyimide was still attached to the glass carrier, these devices demonstrated an average field-effect mobility of 3.15 $cm^2V^{-1}s^{-1}$, a subthreshold slope of 152 mV/dec, and had excellent gate leakage characteristics. After being released from the rigid carrier using PLO, the on-current decreased by 7%, but otherwise the device characteristics showed no appreciable change. The flexible IZO TFTs had an average field-effect mobility of 3.15 $cm^2V^{-1}s^{-1}$ and a subthreshold slope of 152 mV/dec. Importantly, the relatively small change in device characteristics after PLO indicates that the devices did not undergo any changes. This validates the claim that the bulk polyimide remains relatively unheated during PLO.

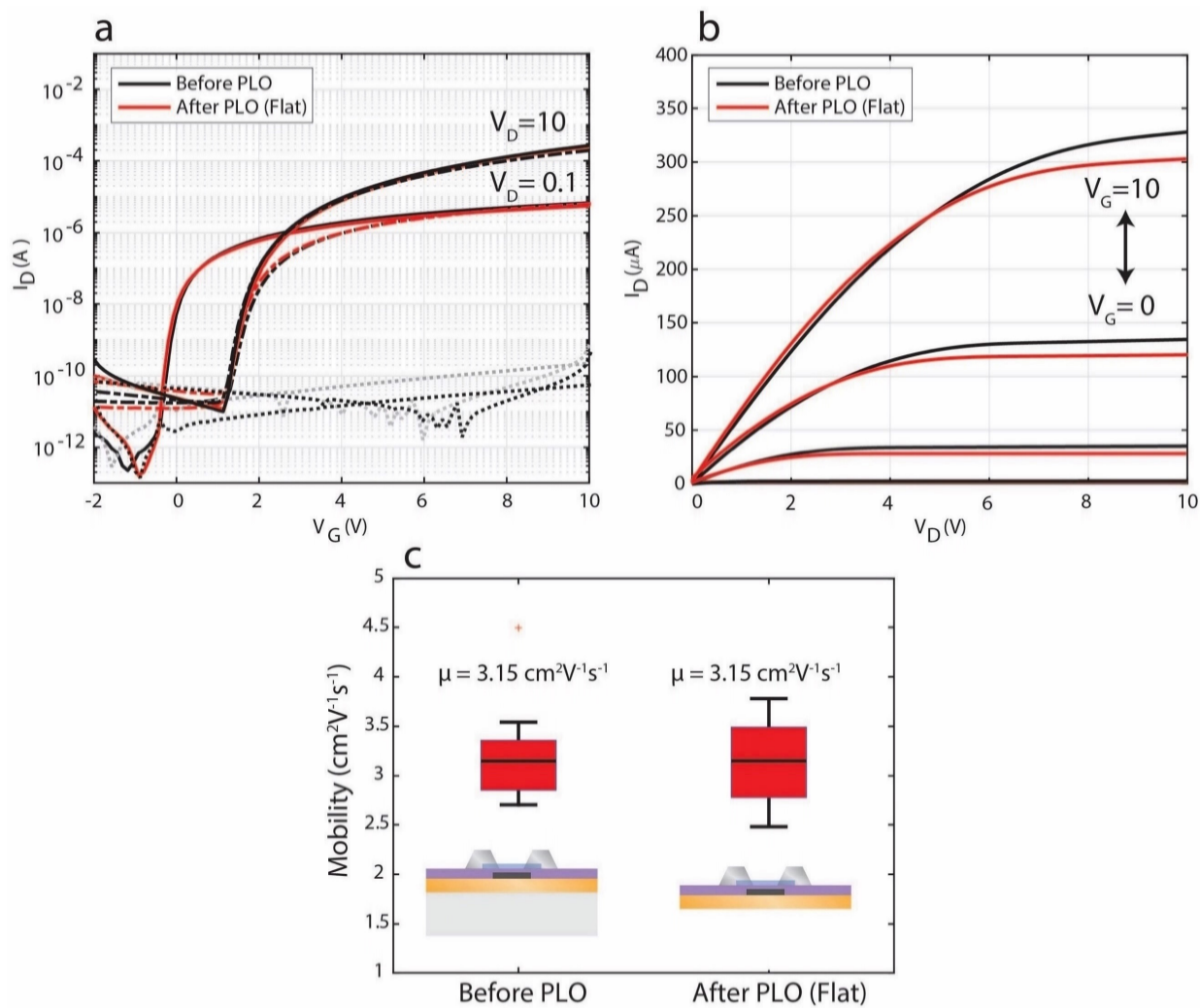

Figure 7: a) $I_D$-$V_G$ transfer curves for devices before PLO and after PLO. b) $I_D$-$V_D$ output curves for devices before PLO and after PLO c) Mobility statistics for IZO TFTs before and after PLO.

After PLO, the mechanical durability of the flexible devices was evaluated by measuring the TFTs while bent to either a 30, 20, or 10 mm radius of curvature. Devices were bent such that the bending axis was parallel to the channel width, placing the channel under tensile stress.

Transfer and output curves measured while the devices were bent are shown in Figure 8, and a summary of device performance is provided in Table 2. Overall, bending produced very little change in the device performance. Devices that were measured while bent to a 30 mm or 20 mm radius of curvature (Figure 8a,d and b,e) have similar performance to both the rigid (unreleased) device and the released flat device, having mobility values of 3.13 $cm^2V^{-1}s^{-1}$ and low gate leakage. The devices measured with a bending radius of 10 mm (Figure 8 c,f) demonstrated a slight reduction in the linear hysteresis and an increase in leakage current. The subthreshold swing value slightly increased from 152 mv/dec to around 170 mv/dec when bending radii of 20 mm and 10 mm were used. This may indicate the formation of defects at the dielectric interface when the sample undergoes bending.

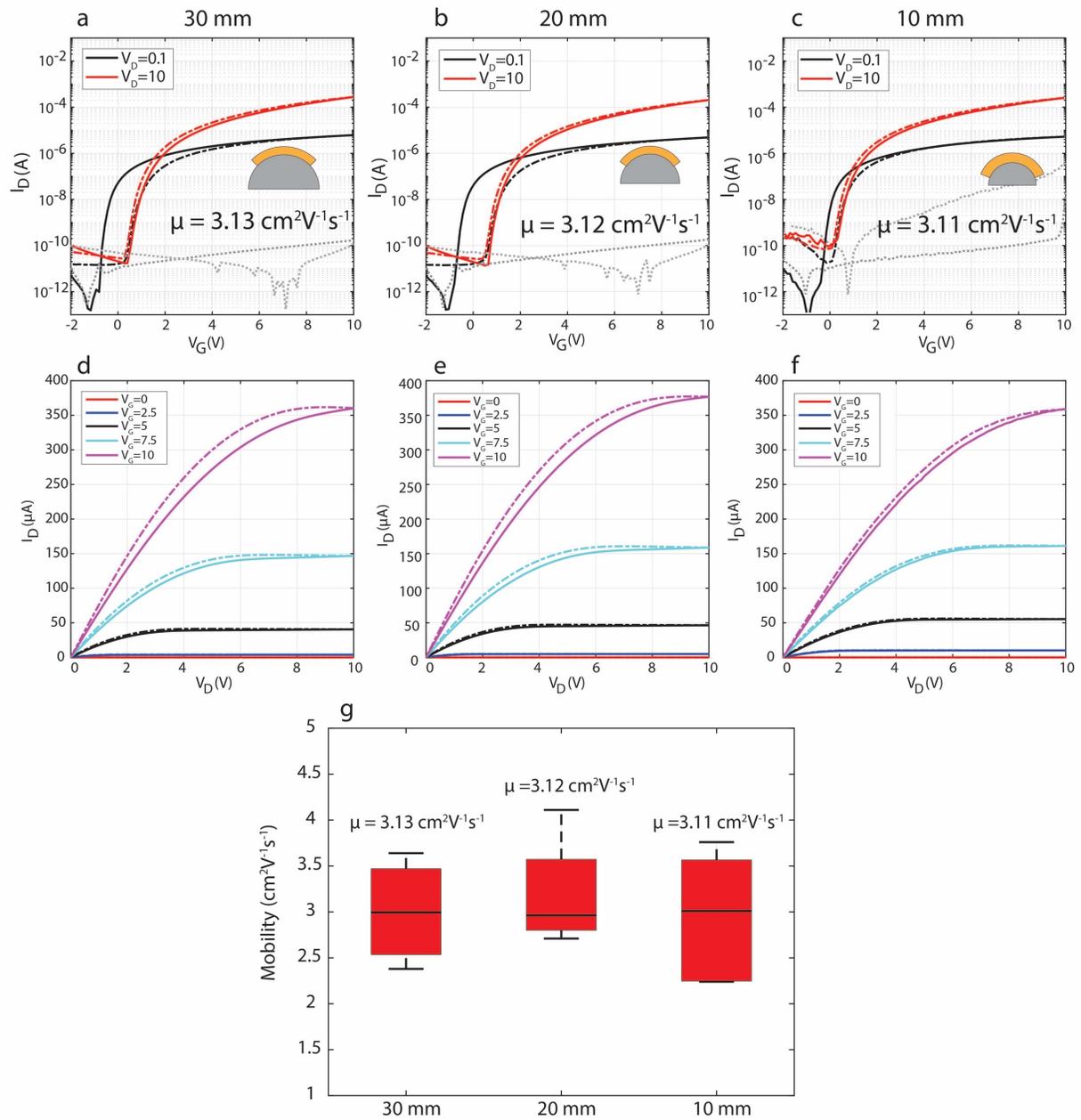

Figure 8: Transfer curves for TFTs under various bending radii: a,d) 30 mm radius; b,e) 20 mm radius; and c,f) 10 mm radius. g) Mobility statistics for devices for various bending radii.

| Radius | $\mu_{FE}$ | Sat. Hysteresis [mV] | SS [mv/dec] | Gate Leakage |
|---|---|---|---|---|
| Before PLO | 3.2 | 130 | 152 | < $10^{-10}$ |
| After PLO (Flat) | 3.15 | 120 | 152 | < $10^{-10}$ |
| 30 mm | 3.0 | 120 | 152 | < $10^{-10}$ |
| 20 mm | 2.96 | 130 | 170 | < $10^{-10}$ |
| 10 mm | 3.01 | 120 | 170 | < $10^{-7}$ |

Table 2: Summary of IZO TFT characteristics before and after PLO, and while bent to a radius of curvature of 30, 20, and 10 mm.

## 2.5 Process Scalability

In this work, the PLO process was successfully used to lift off PI films with IZO TFTs fabricated using 25.4 mm x 25.4 mm Eagle XG glass substrates as the rigid carriers. However, this PulseForge process can also be used to lift off polymeric films from much larger carriers in a matter of seconds. The equipment used here has a single lamp with an irradiation area of 75 mm x 150 mm per flash. The same equipment can irradiate an area of 150 mm x 300 mm when a synchronized flash rate is combined with a motorized table to uniformly process a larger panel. Using proprietary lamp concatenation technology, production versions of this equipment having multiple lamps can irradiate an area of 150 mm x 75*n* mm per flash, where '*n*' is the number of lamps present. For example, a 2-lamp production tool can lift off a 150 mm x 150 mm substrate in one flash. A production version of this equipment with 5 lamps and a 375 mm processing width is shown in Figure S4. This 5-lamp tool was used to demonstrate PI lift-off from a Gen 2 glass carrier (365 mm x 465 mm, Figure 9a), which was accomplished using a series of 4 flashes over 4 seconds. The large-area uniformity of the illumination from these tools is a critical capability that enables PLO to use multiple flashes and multiple passes to achieve rapid, high-yield lift-off. The modeled normalized illumination intensity of the 5-lamp tool for a single flash (375 mm x 150 mm) obtained using a 3-D ray-trace simulation is shown in Figure 9b, with horizontal and lateral uniformity intensity profiles plotted in Figure 9c. As seen in Figure 9b, there is excellent uniformity over the processing area, with less than 5% variability in illumination intensity over the central region. Using the motorized table and synchronized flashing in this manner, we estimated that a polymer film could be lifted off from a 1.5 m x 1.8 m "Gen 6" glass substrate in only ~40 s using 3 passes with a tool containing 8 lamps (Table S1). These results demonstrate that photonic lift-off is a highly scalable and industrially relevant process that can provide efficient, high-throughput polymer lift-off from glass carriers.

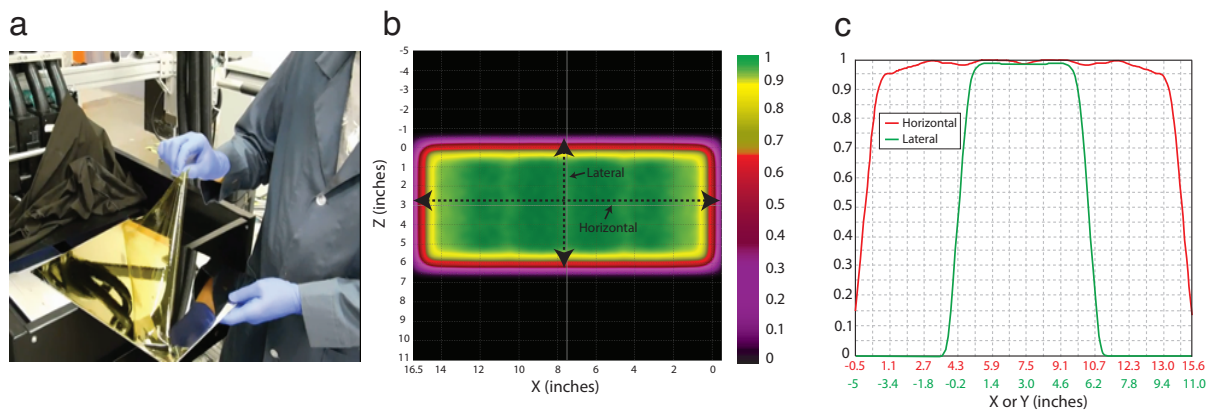

Figure 9: (a) Photonic lift-off of polyimide from a 365 mm x 465 mm Gen 2 glass carrier. (b) Normalized illumination intensity over a single flash area from a 5-lamp PulseForge photonic lift-off tool. (c) Normalized illumination intensity along the horizontal and lateral profiles indicated in (b).

## 3  CONCLUSION

In conclusion, we have demonstrated the fabrication of flexible IZO TFTs on low-CTE polyimide films using photonic lift-off. The TFTs were fabricated using sol-gel IZO and microfabrication techniques (sputtering, lithographic patterning, a wet/dry etching) on a glass carrier substrate coated with a TiW light-absorber layer. The PLO process was completed using a single 950V 150 μs pulse from a flashlamp, which cleanly ablated the PI film from the surface of the light-absorber layer and enabled the PI film containing a TFT array to be successfully lifted off from the LAL. The thermal response as a function of both time and depth into the material stack during PLO were verified with a 3D finite element model, indicating that the PI/LAL interface is heated to 865 ˚C while the surface of the PI containing TFTs does not reach temperatures greater than 118 ˚C. The TFTs were characterized both before and after PLO, showing that they experienced a negligible change in performance after lift-off from the glass substrate and even maintained a constant performance under bending. These TFTs mimic circuitry such as the backplane of a display, thus demonstrating that PLO is compatible with functional thin-film flexible electronics. The metal light-absorber layer allows PLO to be used with a variety of polymer substrates, and the excellent large-area uniformity means it is highly scalable to rapidly release polymer films from substrates over Gen 2 (365 mm x 465 mm) dimension. Photonic lift-off is a robust fabrication platform that can solve many of the dimensional stability issues encountered with polymer substrates and enable high-density flexible electronics.

## 4  METHODS

### 4.1  Light Absorber Layer Deposition

Before depositing the light absorber layer, the starting glass substrate (typically 25.4 mm by 25.4 mm Eagle XG display glass, Corning) was cleaned in a piranha bath consisting of 4 parts sulfuric acid and 1 part hydrogen peroxide. After this, a 250 nm TiW liftoff layer was deposited using magnetron sputtering with the sample platen heated to 250 ˚C. The sputtering power was 250W under 5 sccm of argon. This resulted in a deposition rate of 12.2 nm/min. The light absorber must be sufficiently thick to maximize absorption such that none of the incident optical energy is transmitted into the polyimide film.

## 4.2  Polyimide Film Formation

The PI film was obtained by spin coating PI 2611 (HD Microsystems) over the LAL. In order to ensure good adhesion between the metal LAL and the polyimide, an adhesion promoter VM-651 was used. The adhesion promoter was diluted to a 1:500 ratio of VM-651: DI water and thoroughly stirred prior to spinning. The adhesion promoter was statically dispensed onto the absorber layer ensuring the entire surface was covered and was then allowed to stand for 20 seconds prior to spinning. The sample was spun at 3000 RPM for 40 seconds.

The polyimide precursor (PI 2611, HD Microsystems) was then applied. Prior to spinning the polyimide, it is crucial to ensure that the bottle is brought up to room temperature prior to opening it to avoid residual moisture. If the polyimide is poured into a secondary container for dispensing it must also be degassed to remove any air bubbles. The polyimide precursor was statically dispensed onto the center of the substrate (covering approximately 60%) and allowed to spread for 20 seconds. The sample was then spun at 500 RPM for 5 seconds to ensure sufficient spreading over the substrate and to allow excess PI to spin off. The sample was then slowly ramped up to 2100 RPM (over 10 seconds) where it was held for 35 seconds. After spinning, the sample underwent a 3-step baking process consisting of 2 minutes at 120 ˚C, 2 minutes at 150 ˚C, and 5 minutes at 180 ˚C ensuring that solvents were removed. Three separate hotplates were pre-heated to the specified temperatures, and the sample was transferred between hotplates to avoid the need to ramp up to each setpoint. This spinning and baking step was repeated two more times yielding a 3-layer polyimide film. The sample was then cured in a Sentro Tech programmable tube furnace. The cure consisted of a slow ramp (4 ˚C/min) from 25

˚C to 240 ˚C where it was allowed to dwell for 1 hour. Then the sample was ramped from 240 ˚C at 2 ˚C/min to 380 ˚C where it dwelled for 2 hours. During the ramp from 240 ˚C to 380 ˚C, flowing nitrogen was introduced into the tube. Slow ramp rates are required to avoid rapid degassing a bubble formation in the film. After curing, the sample was then allowed to slowly return to room temperature. This process yielded a polyimide film that was ~30 μm thick with a surface roughness of 0.5 nm$_{RMS}$ (Figure S5). If there is a significant amount of residual back-side contamination from the adhesion promoter and PI, the back side should be cleaned prior to lift-off to prevent PI on the back side from absorbing the incident light pulse and decreasing energy transfer to the LAL.

### 4.3 Thin-film Transistor Fabrication

Bottom gate TFTs structures were fabricated onto the cured polyimide film. A 150 nm molybdenum gate electrode was deposited by DC magnetron sputtering and subsequently patterned using lift-off. The dielectric layer was 42 nm of $HfO_2$ deposited using plasma-enhanced atomic layer deposition (PEALD) at 100˚C. During the PEALD process, the back of the sample was masked off using Kapton tape to ensure the glass remained clean. Immediately prior to depositing in the IZO, the dielectric was cleaned for two minutes using a 200 W, 200 sccm oxygen plasma. This activates the surface of the $HfO_2$ and ensures good wetting of the IZO sol-gel. The IZO sol-gel was then spin-coated onto the dielectric using a static dispense through a 0.3 μm PTFR filter and was then ramped to 3000 RPM and spun for 30 seconds. The sample was transferred to a hotplate and baked at 150˚C for 5 minutes to remove excess solvent. This spin-coating process was repeated to yield a 2-layer film (12 – 15 nm). After spin coating, the device was cured in a Sentro Tech programmable tube furnace. The sample was ramped from room temperature to 350 ˚C over 90 minutes and then held at 350 ˚C for 60 minutes. The sample was allowed to cool gradually to room temperature. After thermal processing, the semiconductor island was patterned using photoresist and then etched in a 5:1 mixture of DI: HCl for 10 seconds. The source/drain electrodes were then formed by lift-off of 100 nm of aluminum deposited by e-beam evaporation. This formed a channel with W/L = 500 µm/60 µm. Finally, a gate contact hole was opened by patterning and etching the $HfO_2$ dielectric in the ion mill.

## 4.4 Photonic Liftoff Tool

Samples were photonically lifted off using a NovaCentrix PulseForge® 1300® or a NovaCentrix PulseForge Invent (IX2-956), each equipped with a 24 mm diameter flashlamp (xenon gas-filled with beam size 150 mm x 75 mm) that emits broadband light (200 nm-1100 nm). The 1300 was equipped with 4 lamp drivers and a 5 kW power supply, while the Invent had 5 lamp drivers and a 6 kW power supply. A National Institute of Standards and Technology (NIST)-traceable bolometer was used to measure the radiant exposure of each pulse used to lift off the samples and calibrate the simulated temperature. The bolometer ensures that consistent processing conditions can be used in each experiment, even across multiple tools. Samples are suspended above the graphite chuck on glass slides, and the sample table height is adjusted 10 mm below the lamp housing. All PLO processing was carried out in an ambient environment.

## 4.5 Device Characterization

TFT device characterization was accomplished using a Keysight B1500 semiconductor device parameter analyzer. Transfer curves and output curves were measured using a forward sweep followed by a reverse sweep of the gate or drain voltage, respectively. Field-effect mobility in the linear (low-field) region was calculated using the transconductance method. The device is operated in the linear region ($V_D$ = 0.1 V), and the linear drain current is differentiated with respect to the gate voltage to obtain the transconductance, $g_m$. The oxide capacitance, $C_{ox}$, was measured using Mo/HfO$_2$/Al capacitors in which the HfO$_2$ was deposited using conditions identical to those used in TFT fabrication. The capacitance was measured with a gate voltage of 5 volts and a frequency of 100 Hz, yielding a capacitance of $C_{ox}$ = 337 nF/cm$^2$. The linear mobility is then extracted using Equation (2).

$$\mu_{FE} = \frac{L g_m}{W C_{ox} V_d} \tag{2}$$

The subthreshold slope was calculated by reciprocating the slope of the $I_D$-$V_G$ transfer curve in the turn-on ($V_{ON} < V_G < V_t$) region. This was plotted against log ($I_D$) and the minimum subthreshold slope was taken from each curve. The TFT turn-on point ($V_{ON}$) is identified as the point on the transfer curve where the current abruptly increases, indicating charge accumulation in the channel. Hysteresis is defined as the difference in $V_{ON}$ between the forward and reverse sweep of the transfer curve measured in saturation. The subthreshold slope was calculated and plotted against the drain current. This method allows to easily exclude any gate current that might contribute to errors in the subthreshold swing.

## 4.6  Finite Element Analysis (FEA) of the PLO process

The 3D FEM model geometry replicated the experimental PLO samples. This consisted of 3 solids representing the glass, LAL, and polyimide. The incident energy was modeled using the deposited beam power module incident on the LAL with a pulse width defined by an analytic equation and energy density determined by bolometry measurements. To capture the abrupt change in applied energy density during the short PLO pulse, an explicit event was triggered at the falling edge of the pulse width. The sample suspension during PLO was modeled by defining a convective heat flux on the external boundaries with a convective heat transfer coefficient of 15 Wm$^{-2}$K$^{-1}$. A swept mesh was applied to all geometries. A mesh distribution was applied to the glass and PI to ensure there was a denser mesh near the LAL/PI and LAL/glass interfaces.

## ACKNOWLEDGEMENTS

This material is based upon work supported by the National Science Foundation under Grant No. ECCS-1710008. Portions of this work were conducted in the Minnesota Nano Center, which is supported by the National Science Foundation through the National Nano Coordinated Infrastructure Network (NNCI) under award number ECCS-1542202. Parts of this work were carried out in the Characterizations Facility, University of Minnesota, which receives partial support from NSF through the MRSEC award number DMR-2011401. The authors would like to acknowledge B. Cote and the Ferry lab at the University of Minnesota for the use of their UV-Vis spectrometer. The authors acknowledge the Minnesota Supercomputing Institute (MSI) at the University of Minnesota for providing resources for COMSOL Multiphysics simulations that contributed to the research results reported within this paper.

## AUTHOR INFORMATION

Affiliations
**Department of Electrical and Computer Engineering, University of Minnesota, MN, USA**
Adam M. Weidling & Sarah L. Swisher

**NovaCentrix, Austin, TX, USA**
Vikram S. Turkani, Vahid Akhavan, & Kurt A. Schroder

## Contributions

A.M.W. designed, fabricated, and characterized devices; developed the 3D simulation model; analyzed simulation and device data; and performed photonic lift-off. V.S.T. and V.A. provided guidance on PLO design and performed sample lift-off. K.A.S and S.L.S supervised the research and contributed to data analysis. All authors assisted with the preparation of the manuscript.

## COMPETING INTERESTS

The authors declare no competing interests.

# Supplementary Information

# Large-Area Photonic Lift-off Process for Flexible Thin-Film Transistors

*Adam M. Weidling[1], Vikram S. Turkani[2], Vahid Akhavan[2], Kurt A. Schroder[2]\*, Sarah L. Swisher[1]\**

**PulseForge Lamp Emission**

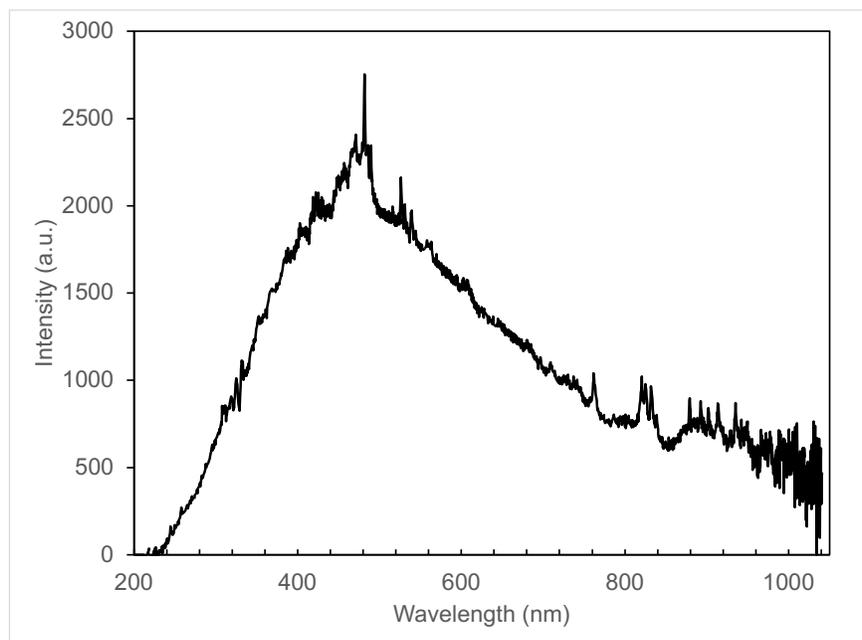

Figure S1: Emission spectrum from the flashlamp in a PulseForge photonic curing system.

# Photonic Lift-Off of Various Polymers

The photonic liftoff process utilizing the light absorber layer has been used to lift off polyimide, clear polyimide, and polyurethane. The light absorber layer creates a robust lift-off platform allowing for various polymers to be released. Depending on the decomposition temperature, the photonic lift-off pulse can be designed by adjusting the capacitor bank voltage and pulse width.

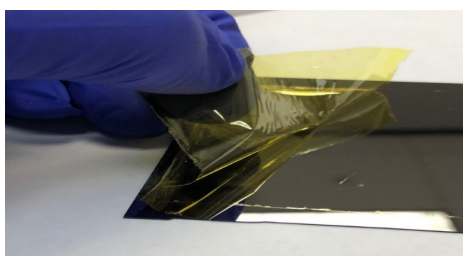
(a)

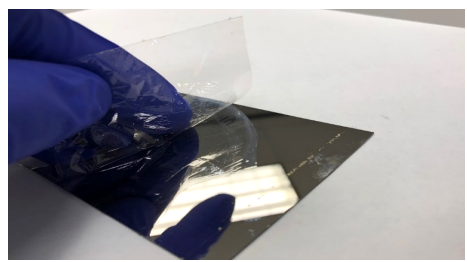
(b)

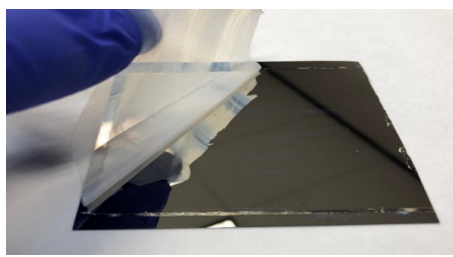
(c)

Figure S2: PLO of (a) polyimide (lift-off at 950V, 150 µs, 4.50 Jcm-2), (b) clear polyimide (lift-off at 950V, 140 µs, 4.13 Jcm-2) and (c) polyurethane (lift-off at 950V, 90 µs, 2.27 Jcm-2).

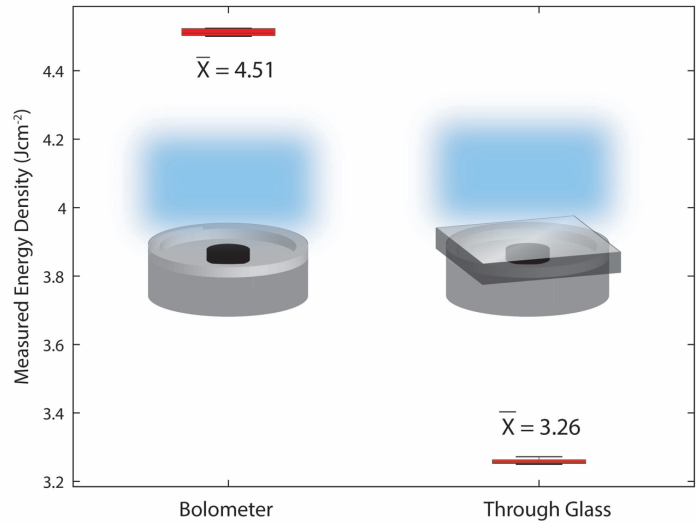

Figure S3: Bolometry measurements at 950 V 150 μs pulse condition for lamp energy incident on a bolometer with and without the Eagle XG glass substrate in place.

# Process Scalability

| Number of Lamps, (n) | Processing Width, n*75 (mm) | Number of Passes | Estimated PLO Process Time (s) |
|---|---|---|---|
| 5 | 375 | 5 | 65 |
| 8 | 600 | 3 | 40 |
| 21 | 1575 | 1 | 15 |

Table S1: Multiple-lamp production equipment with estimated time to process a gen 6 panel (1.5 m x 1.8 m)

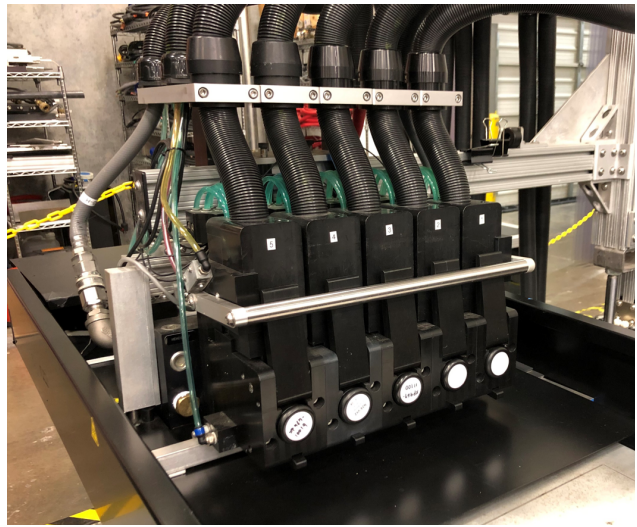

Figure S4: Photograph of the lamp housing for a 5-lamp PLO production tool with 375 mm processing width.

# Polyimide Film Morphology

The surface morphology of the polyimide films was measured using a Digital Instruments NanoScope III scanning probe microscope. The surface scans were taken over an area of 1 μm x 1 μm and were analyzed using Gwyddion software to determine the surface roughness. The resulting polyimide films demonstrated extremely good surface quality having roughness of 0.53 nm$_{RMS}$. This roughness is sufficient to support the development of thin-film devices.

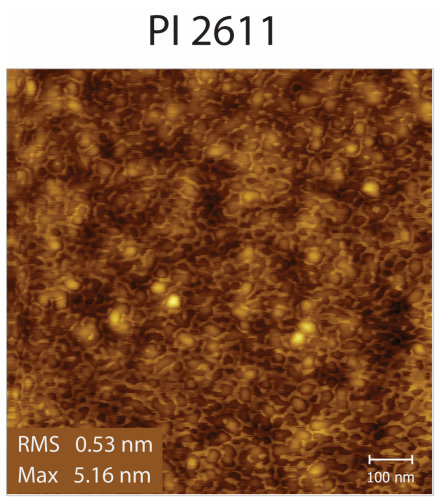

Figure S5: Surface morphology measured with AFM of PI-2611 on a glass carrier.